\documentclass[12pt]{article}
\usepackage{amsmath,amsthm,amsfonts,graphicx,epsfig}
\usepackage[usenames]{color}
\usepackage[active]{srcltx}
\usepackage{amssymb}
\usepackage{graphicx}
 \usepackage{epsfig}
\usepackage{bm}
\usepackage{amsmath}
\usepackage{amsfonts}
\usepackage{ulem}

\def\em{\it}

\newcommand{\ket}[1]{\left| #1 \right\rangle}
\newcommand{\bra}[1]{\left\langle  #1 \right|}
\newcommand{\braket}[2]{\langle #1 | #2 \rangle}

\newcommand{\dbar}{\kern-.1em{\raise.8ex\hbox{ -}}\kern-.6em{d}}
\let\oldhat\hat

\renewcommand{\hat}[1]{\oldhat{\mathbf{#1}}}

\def\half{\mbox{$\frac 1 2$}}

\def\6{\langle }
\def\9{\rangle }
\def \be{\begin{equation}}
\def \ee{\end{equation}}
\def \bea{\begin{eqnarray}}
\def \eea{\end{eqnarray}}
\def \beq{\begin{equation}}
\def \eeq{\end{equation}}

\begin{document}
\title{Distilling entanglement from random cascades with partial
``Which Path" ambiguity}
\author{E. A. Meirom, N. H. Lindner, Y. Berlatzky, E. Poem,\\ N. Akopian,
J. E. Avron and D. Gershoni
\\
%
%
Department of Physics,\\ Technion - Israel Institute of
Technology,\\ Haifa 32000, Israel}
 \maketitle

\begin{abstract}
We develop a framework to calculate the density matrix of a pair
of photons emitted in a decay cascade with {\em partial ``which
path'' ambiguity}. We describe an appropriate entanglement
distillation scheme which works also for certain random cascades.
The qualitative features of the distilled entanglement are
presented in a two dimensional ``phase diagram''. The theory is
applied to the quantum tomography of the decay cascade of a
biexciton in a semiconductor quantum dot. Agreement with
experiment is obtained.
\end{abstract}

%
%
%
%
\section{Introduction}

Two photon cascades with multiple decay paths are candidate
sources of entangled pairs of photons. Practical implementations
of quantum information theory \cite{bennet, gisin} prefer to deal
with qubits that are based only on the photons' states of
polarization \cite{densecoding,teleportation,whiteshor} .
Unfortunately, unless the cascade obeys restrictive symmetry
conditions, the 2-qubit state associated with the polarization of
the photon pair is mixed and has negligible entanglement. As these
symmetry conditions are very hard to achieve, a distillation
procedure is needed in order to obtain entangled polarization
qubits. In this paper we discuss a novel distillation method which
proceeds by spectrally filtering the photons.  The method was
successfully implemented by \cite{nika} in obtaining entangled
polarization qubits from the biexciton cascades in semiconductor
quantum dots \cite{benson}. Our aim is to describe a theory that
allows one to compute the polarization density matrix resulting
from a general decay cascade, with and without distillation.

\begin{figure}[htbp]
    \includegraphics[width=14cm]{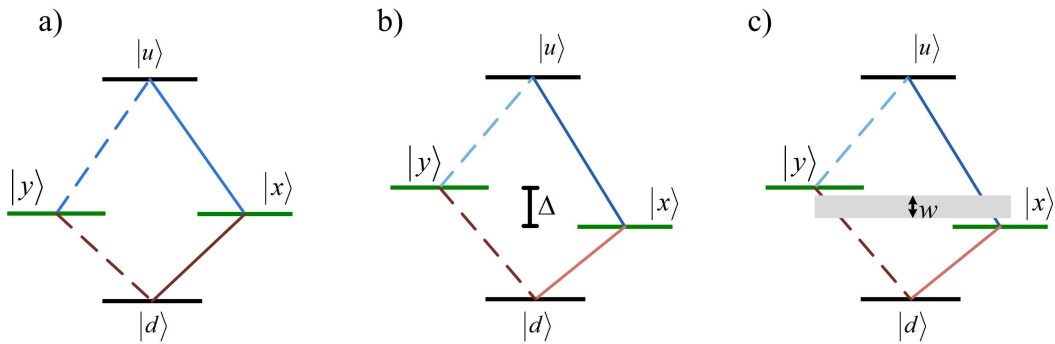}
    \caption{ A decay cascade where the excited state, $\ket{u}$,
    decays to the ground state $\ket{d}$, along two decay paths each
    emitting two photons. The left (right) branch emits two photons that are
    vertically, $y$ (horizontally, $x$) polarized.  In  a) the intermediate level is
    degenerate and the cascade has  perfect ``which path ambiguity''.
    In  b) the  degeneracy of the intermediate
    level is slightly broken by the detuning $\Delta=|E_x-E_y|$. This cascade
    has ``partial which path ambiguity''. c) shows how the
    entanglement can be distilled through spectral filtering
    by a window of width $w$ that erases the
    ``which path'' information    } \label{fig:decay-cascade}
    \end{figure}

The two photon cascades discussed in this paper are illustrated in
Fig.~\ref{fig:decay-cascade}. Each of the two decay paths in the
figure emits a pair of photons with characteristic polarization
and color. In Fig.~\ref{fig:decay-cascade}a  the two decay
channels are distinguished only by their polarization: One channel
gives two horizontally polarized photons and the second channel
gives two vertically polarized photons. In
Fig.~\ref{fig:decay-cascade}b the two decay channels are also
distinguished by the frequencies (colors) of the emitted photons.
When the difference between the photon's frequencies is not too
large (compared with the radiative width of the photons) we call
this ``partial which path ambiguity''  (as the colors of the
photons are not a perfect indicator of the decay path). In
Fig.~\ref{fig:decay-cascade}c the outgoing photons are spectrally
filtered so that only a fraction of the photons, those that do not
distinguish between the decay channels, are collected. These
photons are the ones that have equal probabilities to be emitted
in either channel.


The two photons state, emitted by any one of the cascades in
Fig.~\ref{fig:decay-cascade}, is a pure entangled state. It is
entangled because the quantum decay proceeds simultaneously along
the two decay channels. This, however, does not imply that the
associated pair of qubits, describing the state of polarization,
are entangled.  The state of the qubits is obtained from the
quantum state of the photon field by tracing out all the degrees
of freedom of the two photons (e.g. colors) save the polarization
\cite{Peres}. This state is in general, mixed, and possibly
unentangled in contrast with the two photons state which is pure
and entangled. In fact, partial path ambiguity caused by a
detuning $\Delta$ that is large compared
 with radiative life times---normally the smallest
energy scale in the problem---gives negligible entanglement of the
two qubits. Fortunately, in this case, the entanglement can be
distilled by erasing the ``which path'' information
\cite{eraser,h^4} as indicated in Fig. \ref{fig:decay-cascade}
(c). In fact, by choosing a sufficiently narrow window, one can
distill maximally entangled pairs. The price one pays is that the
probability of finding close to maximally entangled pairs is then
very small.

Decay cascades with ``partial which-path ambiguity''  are
naturally found in the biexciton radiative cascade of
semiconductor quantum dots \cite{takagahara,bayer}. In these
solid-state devices, there is an additional complication in that
the energy levels of the cascade, are (correlated) random
quantities that undergo slow (on the radiative time scale)
fluctuations \cite{empedocles}. These arise from random variations
in the electrostatic potential in the sample. The ensemble of
photons emitted by the cascade is then a mixed state. Entanglement
may or may not not be distilled in the case of general random
cascades with large fluctuations. However, as we shall see, for a
standard model of the random biexciton cascade, distillation works
even when fluctuations are large \cite{nika}.

Our theory allows one to compute the $4\times 4$ density matrix,
$\rho$, of the two photon polarization from the spectral
properties of the cascade. More precisely, we shall see that
$\rho$, is determined by the quantum energies and life times of
the energy levels in the cascade, their distribution, and the
spectral width of the filter. All these quantities can either be
measured or determined by the experimentalist. The theory avoids
modelling the radiating system and we do not need to write a
Hamiltonian for the radiating system. What we do need, instead, is
a ``universal'' form for the photon state generated by a radiating
(dipole) cascade. This 2-photon quantum state depends
parameterically on the energies and lifetimes of the cascade. The
theory  applies irrespective of the nature of the source, be it a
quantum dot, an atom, a molecule etc. It allows us to calculate
the measure of entanglement \cite{h^4} for a given cascade, with
and without distillation. It also allows us to optimize the flux
of entangled pairs.

The paper is organized as follows: In section
\ref{sec:polarization} we describe the polarization density matrix
for cascades with two decay channels. In section
\ref{sec:2photons} we describe the state of the emitted photons
from the radiative cascade in the dipole approximation. We
describe the entanglement distillation in section
\ref{sec:distillation} and the magnitude of the non-diagonal
elements. In section \ref{sec:phase} we discuss the phases of the
non-diagonal elements of the density matrix. In section
\ref{sec:random} we extend the theory to random cascades relevant
to biexciton in quantum dots and in section \ref{sec:experiment}
we compare our theory with the experimental results of Akopian et
al. \cite{nika}.

\section{The polarization density matrix}\label{sec:polarization}

Consider a radiating system, say a quantum dot, inside a
micro-cavity followed by an appropriate optical setup for photon
collection so that the outgoing radiation propagates along the
positive $z$ axis. The polarization of the outgoing photons then
lies in the $xy$ plane. The initial state of the system at time
zero is an excited dot while the photon field is in its vacuum
state, $\ket{u}\otimes \ket{0}$, see Fig.~\ref{fig:decay-cascade}.
For times much longer than the decay time of the dot, $1/\Gamma$,
the dot is in the bottom state and the photon field has a pair of
freely propagating photons, and the quantum state of the dot and
photon filed is $\ket{d}\otimes \ket{\psi}$. Each decay path emits
a pair of photons  with a characteristic polarization: vertical
polarization for the left path and horizontal for the right path
\cite{tejedor}. The state of the freely propagating pair of
photons is then necessarily of the form
 \be   \label{2photons}
    \ket{\psi(t)} =
    \sum_{j=x,y} \int dk_1 dk_2\,\lambda_j\,\alpha_j
    (k_{1} ,k_{2}) e^{ic(|k_1|+|k_2|)t}\,
    {a^{\dag}_{k_1,{j}}a^{\dag}_{k_2,{j}}}
   \ket{0}\ .
    \ee
$\lambda_j$ are the branching ratios for the two decay modes,
$a^{\dag}_{k,{j}}$ is a photon creation operator with wave vector
$k$ and polarization $j$. Since
$a^{\dag}_{k_1,{j}}a^{\dag}_{k_2,j}$ is symmetric in $k_1$ and
$k_2$  only the symmetric part of the functions
$\alpha_j(k_1,k_2)$  contributes to the integral reflecting the
fact that photons are Bosons. We denote by $\alpha^S$ the
symmetrization of $\alpha$ ,i.e.
    \be\label{eq:symmterization}
    \alpha^S(k_1,k_2)=\frac{\alpha(k_1,k_2)+\alpha(k_2,k_1)}2
    \ee

Since the initial state was normalized and the evolution is
unitary, so is the final state
    \be\label{eq:normalization}
    \braket{\psi(t)}{\psi(t)}=\sum_j
    |\lambda_j|^2\braket{\alpha_j^S}{\alpha_j^S}=1,
    \ee

We are interested in the correlations between the polarizations of
two photons.
%
%
This is fully described by the reduced polarization density matrix
whose entries are given by \cite{mandel}
    \beq \label{trace_density}
    \rho_{\mu,\nu;\mu',\nu'}=
    \sum_{k_1,k_2}\bra{\psi}a^{\dag}_{k_2,\nu}a^{\dag}_{k_1,\mu}
    a^{\phantom{\dag}}_{k_1,\nu'}a^{\phantom{\dag}}_{k_2,\mu'}\ket{\psi}
    \eeq

 With $\ket{\psi}$ given by Eq.~\ref{2photons}, one finds for $\rho$
       \be\label{density}
    \rho= \left(%
    \begin{array}{cccc}
    |\lambda_x|^2 & 0 & 0 & \gamma \\
    0 & 0 & 0& 0 \\
    0 & 0 & 0 & 0 \\
    \bar \gamma & 0 & 0 & |\lambda_y|^2 \\
    \end{array}%
    \right),  \quad \gamma=
    \lambda_x^*\lambda_y\braket{\alpha_x^S}{\alpha_y^S}
    \ee
in the basis $\ket{xx}$, $\ket{xy}$, $\ket{yx}$ and $\ket{yy}$
($x$ and $y$ denote the state of polarization). This special form
expresses the fact that the amplitude  for all processes involving
the polarization states $\ket{xy}$ and $\ket{yx}$ vanish. Note
that the matrix has normalized trace, $|\gamma|\le  \half$ and
that the state is mixed for $|\braket{\alpha_x}{\alpha_y}|<1$.

The two qubits are maximally entangled when
$|\lambda_j|^2=|\gamma|=\half$. When $\gamma=0$ the polarization
state is separable and may be thought of as a classical random
source of correlated qubits.  $|\gamma|$ is a measure of the
entanglement known as the negativity \cite{peres-sep,Violating_Bell}, (being the
negative eigenvalue of the partial transposition of $\rho$.)

In the following sections we describe a theory that allows us to
compute $\gamma$ as a function of the spectral properties of
the cascade.

\section{Photons in the dipole approximation}\label{sec:2photons}

To make progress we need to know the functions $\alpha_j$  of
Eq.~(\ref{2photons}).  For this we need to make some assumptions
about the nature of the radiating system. Consider sources that
are small compared with the wavelength of the radiation they emit.
For such sources the dipole approximation applies. We shall
further assume that the interaction between the source and
radiation field is weak so that the rotating wave approximation
applies \cite{scully}. In this setting, which applies to a wide
varieties of radiating systems, the function $\alpha_j$ can be
calculated explicitly. For a radiative cascade with a single
branch this function is given e.g. in \cite{scully,CT}. The case
of two branches is then simply a weighted superposition, as in
Eq.~(\ref{2photons}).

For each branch the function $\alpha_j$ can be expressed in terms
of the spectral properties of the cascade:
$Z_\ell=E_\ell-i\Gamma_\ell, \ \ell=x,y,u$. $E_\ell$ is the energy
of the $\ell$-th state (we chose the ground state to have zero
energy, $E_d=0$) and $\Gamma_\ell$ is its width\footnote{The
common convention \cite{CT} replaces our $\Gamma$ by $\Gamma/2$.}.
For a dipole at the origin one  has \cite{CT}:
    \beq \label{amplitudes}
    \alpha_j (k_{1} ,k_{2})=  A(k_2, Z_j)\,
    A(k_{1}+ k_{2}, Z_u ) \eeq where
    \begin{equation}
\label{A_definition} A(k,Z) = \frac{\sqrt { \Gamma/\pi}}{|k| -Z},
\quad Z=E-i\Gamma
\end{equation}
and we use units where $\hbar=c=1$. This reduces the problem of
computing the entanglement $\gamma$ of Eq.~(\ref{density}) to
computing integrals.


\subsection{The limit of small radiative
width}\label{sec:small-gamma}

In most applications, the radiative widths $\Gamma_\ell$ are the
smallest energy scale in the problem.  This is the case for the
biexciton decay in quantum dot where $\Gamma_\ell\sim 0.8 \mu\,
eV)$, the detuning  $\Delta=|E_x-E_y|\sim 27 \mu\, eV)$ and the
energies of the emitted photons are much larger \cite{ivchenko},
$E_\ell-E_{\ell '}\sim 1.32\, eV)$.

The smallness of $\Gamma$ leads to simplifications in many of the
integrals which can then be evaluated analytically. For example,
$A(k,Z)$ is concentrated near $E$ with a width $\Gamma$, so, in
the limit that $\Gamma$ is small, one makes only a small error by
replacing $|k|$ by $k$. It then follows that, to leading order in
$\Gamma/E$
    \be
    \braket{A}{A}=\int dk |A|^2\approx
    \frac \Gamma \pi \int  \frac{dk}{(k -E)^2+\Gamma^2}= 1
    \ee

In general, as in the case of biexciton decay, the two photons
emitted in each cascade have different colors, namely,
     \be
    \Gamma\ll |(E_u-E_j)- (E_j-E_d)|
    \ee
This distinguishes the two photons which may therefore be treated
as distinct particles and one may forget about the symmetrization,
Eq. (\ref{eq:symmterization}). Mathematically, this follows from
the observation that in computing overlaps, products of the form
            \be
            A(k_1,Z_j)A^*(k_2,Z_j)|A(k_1+k_2,Z_u)|^2
            \ee
are small and can be neglected.

This allows  us to immediately show that the entanglement in a
cascade with partial which path ambiguity is negligible when
$\Delta \gg \Gamma$. This follows from
    \be
    \label{a-alpha}
    {\braket{\alpha_{x}^S}{\alpha_y^S}}\approx {\braket{\alpha_{x}}{\alpha_y}}\approx
    \left( \frac{\Gamma^2}{\Delta^2+\Gamma^2}
    \right)^{\half}\approx \frac \Gamma \Delta
    \ee
The different colors of the emitted photons resolve the ``which
path ambiguity''. This mixes the two qubits and essentially kills
the entanglement.

\section{Entanglement distillation}\label{sec:distillation}

The entanglement can be distilled by selecting those photons which
does not betray the decay path \cite{nika}. Let us denote the
average intermediate states (exciton) energy as $2\bar E=E_x+E_y$.
The first emitted photon do not betray the decay path provided one
only looks at energies near $E_u-\bar E$. Similarly, the second
photon does not betray the decay path provided one only collects
photons with energies $\bar E$.

In practice, the distillation is done by filtering the photons
through a spectral {\em window function}. The photons are detected
only if their energy is either within a window of width $w$
centered at $E_u-\bar E$ or within one centered at $\bar E$. This
is implemented  by a monochromator (an energy filter) which
transmits only a selected part of the emission spectrum
\cite{nika}.

Most photon pairs, are of course, lost in the distillation
process. Roughly, the fraction of photon pairs that are filtered
is of the order $O(w/\Delta)$, as most photons lie in the window of
width $\Delta+O(\Gamma)$. One might worry that, to be effective, the window
must be of the order of the radiative life-time, $w=O(\Gamma)$. If
that was the case, only a very small fraction of the photon pairs
could be distilled. As we shall see, however, this is not the
case. In fact, one may choose $w=O(\Delta)$ so a substantial fraction of the
photons will be distilled while obtaining considerable entanglement. The price one pays for filtering is that
the source is not ``on demand'' but rather a random source of
entangled photons \cite{nika}.

The filtering process is represented in the theory by a projection
operator $W$, which eliminates from a Fock state all photons spare
those whose energy lies within appropriate energy windows
(irrespective of polarization).  Here we are only interested in
the two photon component of the state after filtration. Therefore,
one can effectively express the action of the operator $W$   as
    \be
    W:\alpha_j(k_1,k_2) \to w(k_1)w(k_2)\alpha_j(k_1,k_2)
    \ee
where $w(k)$ is the step function
    \beq
    \label{w_function_defintion}
    w(k)= \left\{
    \begin{array}{ll}  1 \qquad & |k-(E_u-\bar E)|<w/2
    \\ 1 & |k-\bar E|<w/2 \\
                        0  \qquad & {\rm otherwise}

    \end{array}
    \right.
    \eeq
Evidently, $W$ is a projection operator, {\it i.e.} $W^2=W$. The
identity $W=1$ ($w=\infty$) represents no filtering.

\begin{figure}[htbp]
  \centering
  \epsfxsize=.6\textwidth \epsffile{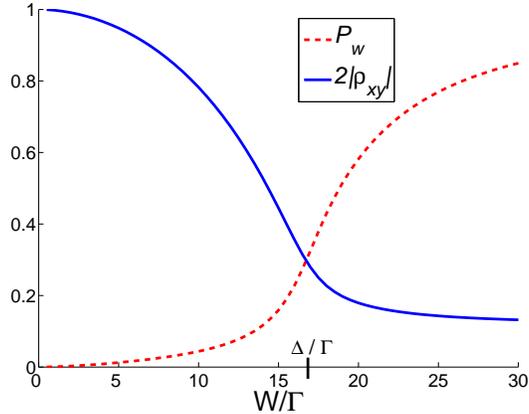}
  \caption{The probability that distillation succeeds is an increasing
  function of the window $w$  shown in the red dotted curve.  The
  entanglement of the distilled pair is a decreasing function of
  $w$ shown in the blue curve. The pair is maximally entangled at
  $w=0$. The plot is drawn for parameter values corresponding
  to biexciton decay where $\Delta/\Gamma\approx 17$. }
  \label{fig:norm_mag}
\end{figure}

The distillation succeeds with probability
$p_W={\bra{\psi}W\ket{\psi}}$ and produces the
(normalized) filtered state
    \be\label{filtered-state}
    \ket{\psi^f}= \frac{W\ket{\psi }}{\sqrt{p_W}}
    \ee
The filtered, or distilled, density matrix can be computed from
the distilled state. In particular, for the entanglement, as
measured by $\gamma$ of Eq.~(\ref{density}) we find
    \be\label{eq:distilation}
    \gamma_d=\lambda^*_x\lambda_y
    \frac{\bra{\alpha_x^S}W\ket{\alpha_y^S}}{p_W}, \quad p_W=
    \sum_{j=x,y} |\lambda_j|^2\bra{\alpha_j^S}W\ket{\alpha_j^S}
    \ee
This reduces the problem to computing integrals, where we account
for $W$ by summing only  the appropriate wavevectors.
Fig.~\ref{fig:norm_mag} shows the probability to detect a pair of
photons and $\gamma_d$ of the distilled state, as function of the
width of the spectral window $w$. To plot the figure we use
parameter values corresponding to biexciton decay in a quantum
dot. As one expects, the entanglement is a decreasing function of
$w$, (for a window of zero width one gets a maximally entangled
state). On the other hand, the probability that the detection
succeeds is, of course, an increasing function of the width.

\begin{figure}[htbp]
  \centering
  \includegraphics[width=8cm]{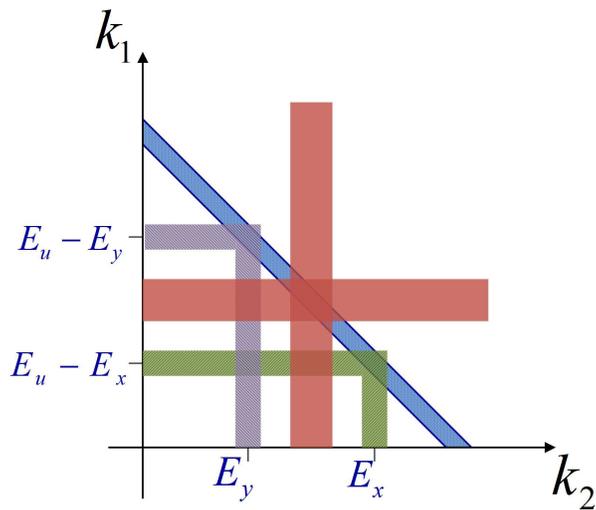}
  \caption{The wave function $\alpha_x$ is large near the intersection
  of the diagonal strip, expressing total energy conservation, with the
  line which represents the $x$ decay path where the second
  photons has energy $E_x$. Similarly, the
  function $\alpha_y$ is large near the intersection of the  diagonal
  strip
  with the  line representing the $y$ decay path where the
  second photon has energy $E_y$.
  The cross represents a filter of narrow width.
  The filter collects photons that are contained in the
  intersection of the cross.
  }
  \label{fig:functions}
\end{figure}

The qualitative behavior of entanglement distillation can be
gleaned by inspection of Fig. \ref{fig:functions}. The function
$\alpha_x$ is concentrated in a small neighborhood of size
$O(\Gamma)$ near the point of intersection of the green and blue
curve. Similarly, $\alpha_y$ is concentrated near the intersection
of the purple and blue curve. For example, the fact that the
entanglement without distillation is small, Eq.~(\ref{a-alpha}),
follows from the little overlap of $\braket{\alpha_x}{\alpha_y} $
each of which is concentrated near a different point. Due to
distillation, only amplitudes contained in the intersection of red squares are
collected. This does two things. It {\em decreases} the numerator
in Eq.~(\ref{eq:distilation}) which is bad. However, it also
decreases the denominator which is good. This decrease is much
more significant and consequently the entanglement increases.

Perhaps the most interesting things one learns form Fig.
\ref{fig:functions} is how wide does a window have to be to betray
the ``which path information''. This happens when the window
contains the points of intersections,  either red with purple, or
red with green. If the window does not contain these points the
path is not betrayed. Since the points have a small size,
$O(\Gamma)$, this implies that the size of the optimal window is
of the scale of the detuning, $w=\Delta-O(\Gamma)$. Because this
window is not small, the probability that the distillation
succeeds is not very small either.




\section{The phase problem}\label{sec:phase}

From the perspective of quantum information theory the phases in
the density matrix are gauge dependent quantities (as they are not
invariant under local unitary operations \cite{locc}). Even if
Alice and Bob fix the projectors $\ket{0}\bra{0}$ and
$\ket{1}\bra{1}$, there is still a freedom to choose the phases of
the states
    \be
    \ket{a}\otimes \ket{b}\to
    e^{i\phi_a}e^{i\varphi_{b}}\ket{a}\otimes\ket{b}, \quad
    a,b\in\{0,1\}.
    \ee
We refer to this as gauge freedom. Such a transformation will
change the phases of the non-diagonal entries of the density matrix
    \be\label{eq:ambiguity}
    \rho_{ab,cd}\to
    e^{i(\phi_a+\varphi_b-\phi_c-\varphi_d)}\rho_{ab,cd}, \quad
    a,b,c,d\in\{0,1\}
    \ee
    in the computational basis. Any reasonable measure of
entanglement,  and in particular $|\gamma|$ of
Eq.~(\ref{density}), is  clearly independent of the choice of
gauge.

Quantum tomography \cite{kwiat} is an algorithm to convert 16
measurements to the 16 (complex) entries of the density matrix
$\rho$ (describing the ensemble) \cite{quorum}. This means that
any quantum tomography algorithm must fix both the projectors
representing the ``computational'' basis and fix the gauge.

In the context of photon polarization the canonical choice (which
we used throughout this paper) of the ``computational'' basis is
the projectors associated with the $x$ and $y$ linear
polarizations. The remaining  gauge freedom is
    \be
    a_{kj}\to e^{i\phi_j} a_{kj}, \quad  a_{kj}^\dagger\to e^{-i\phi_j}
    a_{kj}^\dagger \qquad j=x,y
    \ee
for two orthogonal polarizations $j$. Fixing the right circular
polarization by
    \be
    a_{k,R}^\dagger = \frac{a_{kx}^\dagger  + i
    a_{ky}^\dagger}{\sqrt 2}
    \ee
fixes the gauge since
 \be
    a_{kR}\to e^{i\phi_R} a_{kR}, \quad  a_{kR}^\dagger\to e^{-i\phi_R}
    a_{kR}^\dagger\, ,
 \ee
requires that all the  $\phi$'s are the same. This then fixes the
phase of $\gamma$.

The phase of $\gamma$, which was measured in
\cite{nika,edamatsu,Shields-njp} have, so far, not been explained
by a theoretical model. In the following, we calculate this phase
and describe the physical information that is encoded in it.

Eq.~(\ref{eq:distilation}) determines $\gamma$ in terms of the
product of the branching ratios $\lambda_x\lambda^*_y$ and the
overlap $\bra{\alpha_x^S}W\ket{\alpha_y^S}$. In the next section, we shall show that for a decay
cascade with partial which path ambiguity and time-reversal
invariance, $\lambda_x\lambda^*_y>0$. It then follows that the phase of $\gamma$ is fully determined by the phase of
$\bra{\alpha_x^S}W\ket{\alpha_y^S}$.

\subsection{Branching ratios} \label{time-reversal}


 The branching amplitudes
$\lambda_j$ of Eq. (\ref{2photons}) are proportional to the
appropriate dipole matrix elements
%
\begin{eqnarray}
\lambda_x &=& \zeta \bra{u} X \ket{j} \bra{j} X \ket{d} \nonumber \\
\lambda_y &=& \zeta \bra{u} Y \ket{j} \bra{j} Y \ket{d}
\end{eqnarray}

where $\zeta$ is a overall normalization constant and $X$
is the x component of the (possibly multi-electron) position
operator and similarly $Y$ is the y component of the position operator. It follows that
    \be\label{lambda-lambda}
    \lambda_x^*\lambda_y=|\zeta |^2\, \bra{d} X \ket{x}\bra{x}
    X \ket{u}
    \bra{u} Y \ket{y} \bra{y} Y \ket{d}
    \ee
Observe first, that this quantity is independent of the gauge
choice of the states $\ket{\ell}$ of the source, as every ket is
paired with the corresponding bra. We shall now show that in the
case that all the states $\ket{\ell}$ are non-degenerate, there is
a choice of gauge so that each matrix element in the product is
real.

Let $T$ denote the antiunitary operator associated with time
reversal \cite{sachs,messiach}, i.e.
    \be
    \braket{T\ell}{Tk}=\braket{k}{\ell}
    \ee
In the case that all states $\ket{\ell}$ are non-degenerate
$T\ket{\ell}=e^{i\beta_\ell}\ket{\ell}$. By changing the gauge to
$\ket{\ell}\to e^{i\beta_\ell/2}\ket{\ell}$, one  sees that
$\ket{\ell }$ may be chosen so that $T\ket{\ell }=\ket{\ell}$. The
position operator is evidently even under time reversal e.g.
$T^* X T=X$. Plugging this in the definition of
the dipole matrix elements we see that
    \be
     \bra{\ell}X\ket{k}= \bra{\ell}T^{*} X T\ket{k}=
     \bra{T\ell}X\ket{Tk}
    =\bra{k}X\ket{\ell}
    \ee

We have therefore shown that $\lambda_x^*\lambda_y$ is a real
number.  We shall now show that under rather weak continuity
assumptions, it must actually be positive. $\lambda_x^*\lambda_y$
is a function of the spectral properties of the cascade, and in
particular, it is a function of the detuning $\Delta$. It has the
same sign as $\Delta$ varies so long as the two decay paths are
indeed {\em effective} (none of the branching ratios, $\lambda_j$,
vanishes). It is therefore enough to determine the sign at a
single point. We shall now give a symmetry argument that at
$\Delta=0$ one has $\lambda_x^*\lambda_y>0$.


Assume that the degeneracy  $\Delta=0$ is a consequence of
(possibly approximate) rotational symmetry in the x-y plane of the
radiating system, (this is the case in quantum dots). Since the
sign of the product of dipole matrix elements changes continuously
as the Hamiltonian is deformed, we may compute the sign for the
case where the rotational symmetry is exact. In this case, as the
initial, non-degenerate, state $\ket{u}$ must be a state of
angular momentum 0 about the z-axis. Since angular momentum is
conserved the final two photon state must also be a state of zero
angular momentum about the z-axis.

In this case, perfect which path ambiguity and zero angular
momentum imply that the state of the outgoing photons is
    \be   \label{2photons-d}
    \ket{\psi(t)} =
    \int \frac{dk_1 dk_2}{\sqrt{2}}\,\alpha
    (k_{1} ,k_{2}) e^{i(|k_1|+|k_2|)t}\,\left(
     a^{\dag}_{k_1,R}a^{\dag}_{k_2,L}+
     a^{\dag}_{k_1,L}a^{\dag}_{k_2,R}\right)
   \ket{0},
    \ee
By comparing this with Eq.~\ref{2photons} one easily sees that
this state implies $\lambda_j=1/\sqrt{2}$. Hence, $\gamma=\half$
in Eq.~(\ref{density}),  which determines the sign of the product
$\lambda_x^*\lambda_y>0$.

\subsection{The role of the complex pole} \label{subsec:pole}

It follows from the analysis above that the phase of $\gamma$ is
the same as the phase of $\bra{\alpha_x}W\ket{\alpha_y}$. The
latter is determined by a two-dimensional integration of the
function
 \begin{equation}
 \label{eq:amplitude}
  w(k_1)w(k_2)\
\Big|A(k_1+k_2,Z_u)\Big|^2\,  A(k_2,Z_j) \bar A(k_2, Z_{j'})
  \end{equation}
The first three factors are positive, and weigh the integrand. The
third factor  may be interpreted as guaranteeing approximate
conservation of total energy since
    \be
    \lim_{\Gamma\to 0} \Big|A(k,Z)\Big|^2=\delta(|k|-E)
    \label{eq:energy conserv}
     \ee This means that to leading order in
$\Gamma$ the matrix elements of $\rho$ are determined by a
one-dimensional integral over $k$ of the function
 \begin{equation}
 \label{eq:r1r2}
  w(E_u-k)w(k)\  A(k,Z_j) \bar A(k, Z_{j'})
  \end{equation}
The phase of $\gamma$ is governed by the phase of $A(k,Z_x) A^*(k,
Z_y)$ which is represented graphically in Fig.~4

\subsubsection{Strong filtering: } Suppose the filtering window $W$
is very narrow with width $\Gamma<w\ll \Delta$.  The window
restricts the domain of integration to a very narrow region. The
detection probability is small and scales linearly with the
window's width $p_W=O(\frac{w\Gamma}{\Delta^2})$. The
phase of $\gamma$  is
    $\pi-4\Gamma/\Delta$ and the magnitude is
approximately.
    \beq
    \gamma=-\half+O\left(\frac{\Gamma}{\Delta}\right)
    \eeq
    The state is close to a maximally entangled state.
This gives the upper left triangle of Fig.~\ref{fig:phase} and the
left part of Fig.~\ref{fig:norm_mag}.

\subsubsection{No filtering: }
No filtering corresponds to $W=1$ and a width $w=\infty$. Exact
degeneracy, $\Delta=0$, gives a maximally entangled state,
$\gamma=1/2$. (The two arrows in Fig.~4 are complex conjugates.)

When $\Delta \gg \Gamma$ the integrals are dominated by the
neighborhood of the poles at $k=E_x,E_y$. The off-diagonal element
is almost purely imaginary and $\gamma=O\left( i\frac \Gamma {\bar
\Delta}\right)$. This accounts for the lower right hand triangle
of Fig. \ref{fig:phase} and the right hand part of
Fig.~\ref{fig:norm_mag}.

\begin{figure}[htbp]
\centering
\begin{tabular}{lc}
(a) \\
\epsfxsize=.3\textwidth \centerline{\epsffile{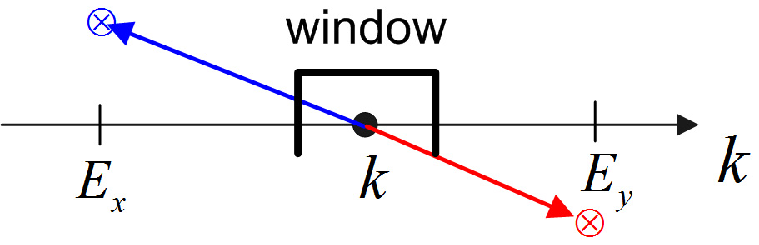}} \\
(b) \\
\epsfxsize=.3\textwidth \centerline{\epsffile{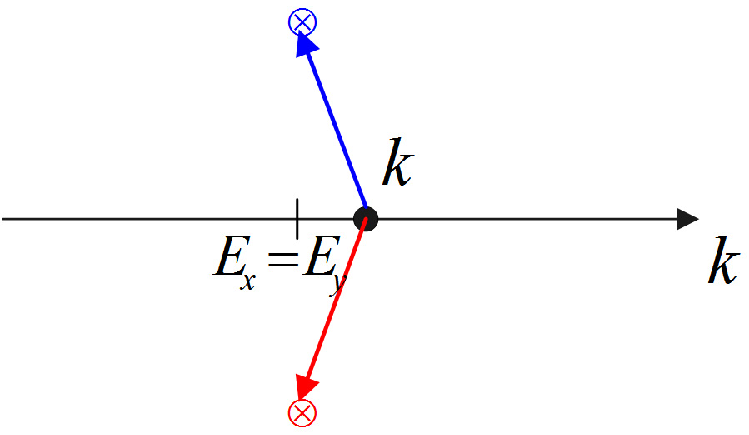}} \\
(c)\\
\epsfxsize=.35\textwidth \centerline{\epsffile{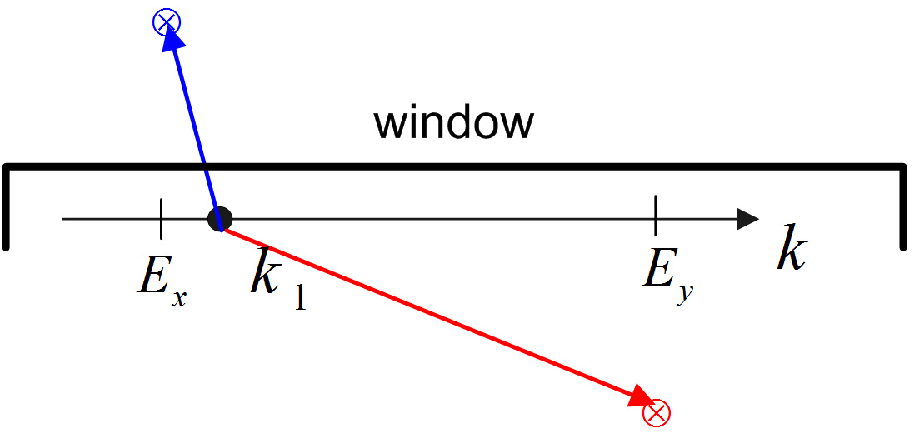}}
\end{tabular}
\caption{The phase is determined by the product of the two complex
numbers $(k-Z_x)(k-Z_y^*)$. The numbers are represented as arrows
pointing from $k$ to the location of the complex energies
$Z_{\varepsilon}$ (blue arrow for $Z_x$ and red arrow for $Z_y$).
The location of $k$ is restricted by the window function $W$. In
figure (a) the levels are detuned and the spectral window is
smaller than the detuning. In figure (b) the levels are degenerate
and in figure (c) the levels are detuned and the window is larger
than the detuning.} \label{fig_arrows}
\end{figure}


 \begin{figure}[htbp]
 \epsfxsize=.45\textwidth \centerline{\epsffile{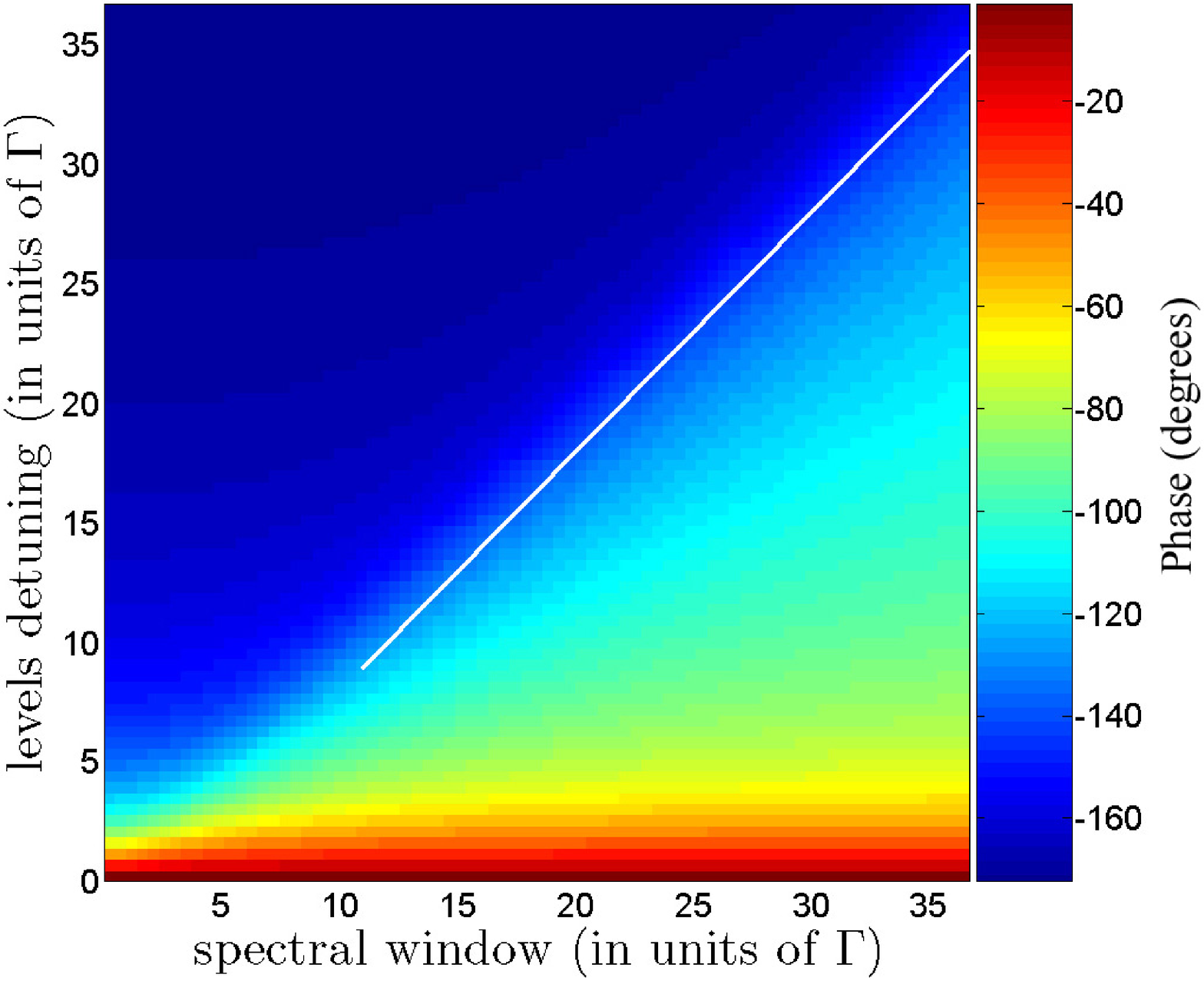}}
\caption{The phase of $\gamma$ as a function of the (centered)
normalized spectral window width $w/\Gamma$ and the detuning
$\Delta/\Gamma$ for parameters appropriate to biexciton in a
 quantum dot.
The triangular reagin  above the diagonal represents the situation
of a narrow filter where $\gamma\sim -1/2$. The triangular region
below the diagonal is where the window is large, $\gamma$ is small
and essentially purely imaginary. At the bottom of the figure the
detuning is small and $\gamma \sim 1/2$.
   }
 \label{fig:phase}
 \end{figure}

\section{Random cascades}\label{sec:random}
Radiative cascades with partial which path ambiguity are found
naturally in semiconductor quantum dots where $\ket{u}$ is the
ground state of a bound state of a pair of two electrons and two
holes (a biexciton). The states $|x\rangle$ and $|y\rangle$ are
the ground and first excited state of the bright exciton (a bound
electron-hole pair). The state $|d\9$ describes an empty quantum
dot.  In this case the energies and states slowly fluctuate due to
electrostatic transients in the semiconductor hosting the quantum
dots. In typical cases \cite{nika}, the fluctuations are large
(comparable to $\Delta$)  and slow (the time scale is much longer
than $1/\Gamma$).
\begin{figure}[htbp]
 \centering
    \includegraphics[width=10cm]{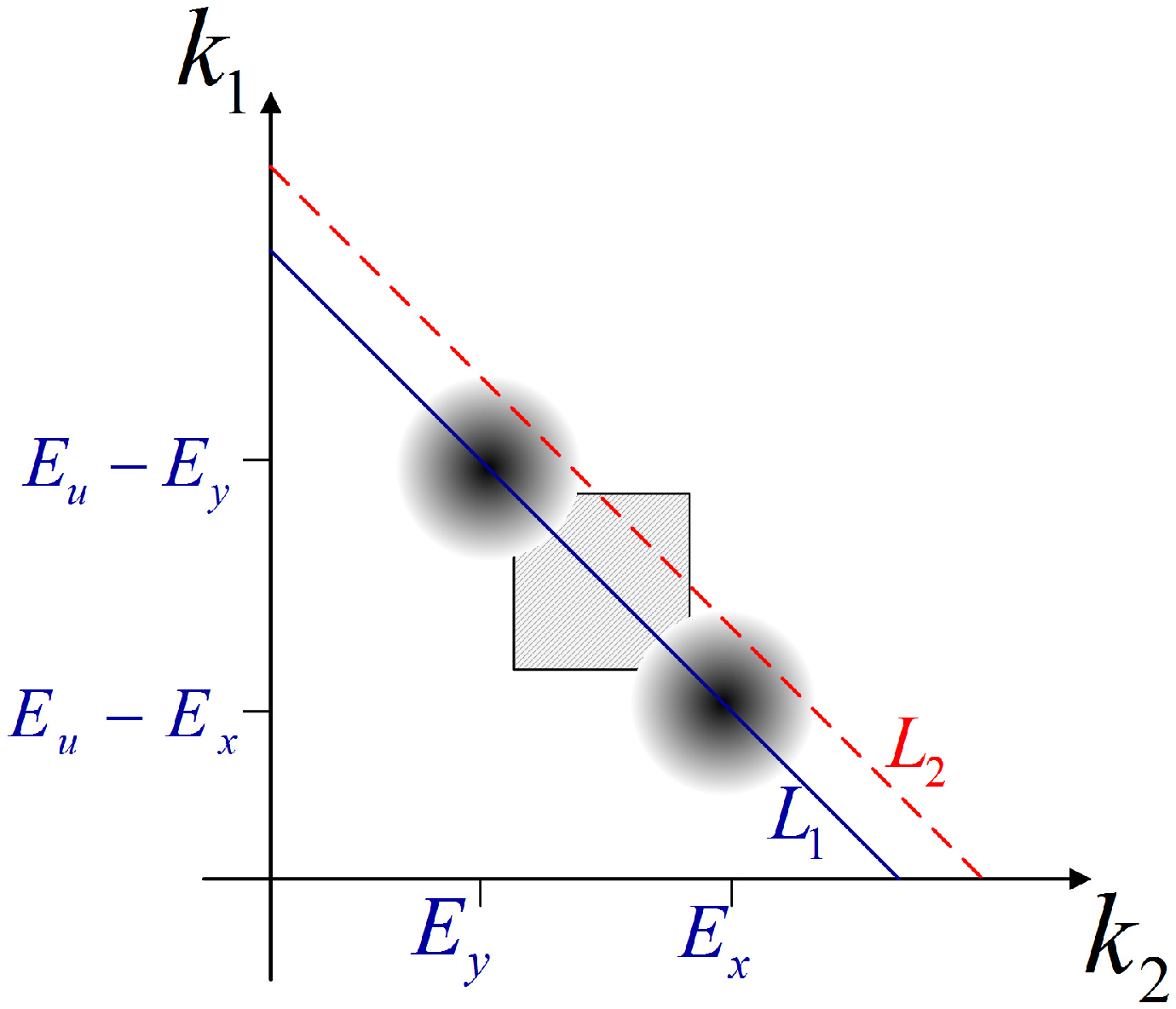}
 \caption { The gray shapes represent the fluctuations of the cascade.
 The lines diagonal represent conservation of total energy and correlate the
 two gray shapes. They fluctuate as well. The rectangle represents the
 filtering. In  a) some of the events penetrate the filter and thus
 betray the which path ambiguity. In b) the fluctuations do not betray the which path
ambiguity and entanglement can be filtered. }
\label{fig:random-bad}
\end{figure}


One can model the situation by letting the spectral properties of
the cascades, namely $Z_\ell(s)$, be appropriate functions of a
random variables $s$, with measure $dP(s)$. The two-photon state
of Eq.~(\ref{2photons}), $\ket{\psi(s)}$ is then a random
variable.

The photon field $\ket{\psi(s)}$ of Eq.~(\ref{2photons}), depends
on $s$ through the fluctuating complex energies, $Z_\ell$, of
Eq.~(\ref{amplitudes}). The 2 photon state emitted by the dot is
then described not by a pure state but rather by a density matrix
    \be
    \rho_r=\int dP(s) \ket{\psi(s)}\bra{\psi(s)}.
    \ee
The probability to distill a state describing a specific random event is
    \be
    p(s)=\bra{\psi(s)}W\ket{\psi(s)}.
    \ee
Therefore, the probability to distill a photon pair is given by
\be p(W|\rho_r)=\int p(s)P(s)ds \ee
 Similarly,  the value of the distilled $\gamma_d$ is given by
 averaging over $s$

    \be
   \gamma_d= \frac{\int dP(s)\, p(s)\, \gamma_d(s)}
    {\int dP(s) p(s) }
    \ee
where $\gamma(s)$ is given by substituting $\ket{\psi(s)}$ in
Eq.~(\ref{eq:distilation}).

In general, large spectral fluctuations can potentially destroy
the distillation based on fixed spectral windows. This, for
example, is the situation when the values of $E_u,E_x$ and $E_y$
are independent random variables, as illustrated in
Fig.~\ref{fig:random-bad}(a). In the figure, the amplitudes which
betray the which path ambiguity penetrate the filter. When the
fluctuations are smaller than $\Delta$ one can remedy this by
choosing a sufficiently small spectral window. This situation is
illustrated in Fig.~\ref{fig:random-bad}(a). However, when the
fluctuations are larger than $\Delta$ distillation  becomes
impossible.

A scenario which allows for distillation also when the
fluctuations are larger than $\Delta$ is illustrated in
Fig.~\ref{fig:random-bad}(b).

At first, one may think that the second ``good'' scenario is
contrived and would not naturally occur.  In fact, this is not the
case and this scenario is the one that describes biexciton drift.
The reason is that the energies $E_u$, $E_y$ and $E_x$ are {\em
not} independent random variables, as in Fig.
\ref{fig:random-bad}(a), but rather {\em dependent} random variables.
This is because for a biexciton, $E_u=E_x+E_y-B$ where $B$ is the
biexciton binding energy \cite{bimberg} which is typically more than two
orders of magnitude smaller than the exciton energy and its
dependence on $s$ can be safely ignored.  A model for a
fluctuating spectral diagram is then
    \be
    E_x\to E_x+s, \quad E_y\to E_y+s,\quad E_u\to E_u+2s
    \ee
This indeed leads to a scenario like the one illustrated in
Fig~\ref{fig:random-bad}(b).

We note that the biexciton drift described above has only little
effect on the calculation of $\gamma_d$  described in
section~\ref{sec:phase} and Fig~\ref{fig:phase}. This results from
the rapid decrease of the probability of detection $p(s)$ from its
maximal value at $s=0$. This can be seen in Fig.~\ref{fig:shift
prob} which plots $p(s)dP(s)$ for finite value of $\Gamma$.

\begin{figure}[htbp]
\epsfxsize=.4\textwidth \centerline{\epsffile{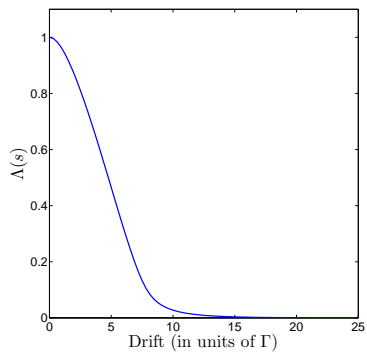}}
\label{fig:shift prob} \caption{The relative ``weight" of the
density matrix $\rho(s)$. The weight is given by $\Lambda(s)=P(s)
Tr (W\rho(s))$. The plot is renormalized to yield $\Lambda(0)=1$.
The plot is obtained with the experimental values as in
Sec.~\ref{sec:experiment}.}
\end{figure}

\section{Comparison with experiment}
\label{sec:experiment}

We now turn to compare the theoretical calculation with the
experimental data as described by Akopian et al. \cite{nika}

The parameters used in the theory were measured
independently\footnote{We are using the half-width at half maximum
(HWHM) convention, while in \cite{nika,CT} the radiative width is
given according to the FWHM convention.}: $ \Gamma_x \approx
\Gamma_y = 0.8 \pm 0.2 [\mu eV],\ \Gamma_u \approx 2\Gamma_x$ and
$E_x \approx E_y=1.28 [eV],\ E_u= 2.55 [eV]$, $\Delta= 27 \pm 3
[\mu eV]$.  The window that was used in the experiment was of
width $w=25 \pm 10 [\mu eV]$.

The distribution $P(s)$ of the spectral shift was evaluated from
the measured spectral lines. It was rather wide, with full width
middle height of about $50  [\mu eV]$. With the values listed
above the probability of detection $Tr (W\rho(s))$ falls much
faster then the distribution $P(s)$ as a function of $|s|$, to
half its value at $|s|\sim 5 \Gamma \sim 8 [\mu eV]$.

The numerically calculated contribution to the filtered state
(\textit{i.e.} probability of detection times probability
distribution for $s$) as a function of the spectral drift $s$ for
the above parameters is displayed in Fig.~\ref{fig:shift prob}.

\begin {figure}[htbp]
\epsfxsize=.4\textwidth
 \centerline{\epsffile{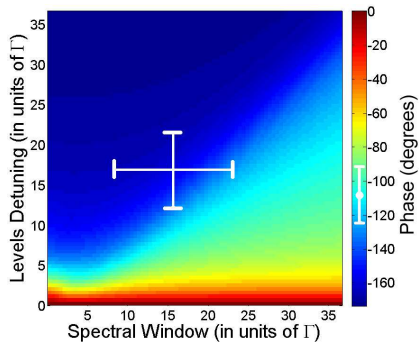}}
\caption{ Comparison between the experimental results of Akopian
et al. and the theory. The graph shows a theoretical calculation
of the phase as a function of the window, $w$ and the detuning,
$\Delta$, both in units of $\Gamma$. The calculation uses the
experimentally measured parameters $w$, $\Delta$, $\Gamma$ and
$s_0$. The uncertanties in these parameters result in an area
(rather than a point) indicated by the error bars. The possible
theoretical values of the phase are in the area which is bounded
by these error bars. This values are compared to the
experimentally measured phase and error, which is represented in
the color bar to the left.}

\label{fig:comparison}
\end {figure}

When we come to compare the theory with the experimental results,
we must take into account the measurement error of the tomography,
as well as the errors on the parameters $\Delta, \Gamma$,$s$ and
$w$ (the QD or model parameters). These are displayed in
Fig.~\ref{fig:comparison}. The measured phase in the experiment
was $-110^{\circ} \pm 17^{\circ}$  where we have taken into
account the effect of the beam splitter. The beam splitter induces
a phase shift of $180^{\circ}$ between the X and Y polarizations
of the reflected photon only (this was ignored in Akopian et al.
\cite{nika}, where the phase was reported $70^{\circ}$). This is
compared to the theoretical result $-160^{\circ} \pm 45^{\circ}$,
which shows a reasonable fit with the experiment.

\section{Conclusion}

We described a framework to calculate the density matrix of a pair
of photons emitted in a decay cascade with {\em partial which path
ambiguity}, encoded in the energies of the emitted photons. We
showed that one can distill the entanglement by selecting only the
photons possessing ''which path" ambiguity and discuss how this
distillation by spectral filtering affect the phase of the
non-diagonal elements of the two photon density matrix. We showed
that spectral filtering is quite robust and protected from
fluctuations in the level's energies as long as these fluctuations
are correlated. Our calculations, quantitatively describe
measurements performed on semiconductor quantum dots.


\section*{ACKNOWLEDGMENT} This work is supported by the Israeli
Science Foundation, the Russel Berry Institute for Nano Technology
and the Fund for Promotion of the Research at the Technion.

\bibliographystyle{unsrt}
\bibliography{amalgam_final}

\end{document}